{
\def\a{\alpha}
\def\b{\beta}
\documentstyle[preprint,aps]{revtex}

\begin{document}
\draft
\preprint{}

\title{Kinetic Inductance of Josephson Junction 
	Arrays: Dynamic and Equilibrium Calculations}
\author{Wenbin Yu and D. Stroud\\
{\it Department of Physics, The Ohio State University, Columbus, 
     OH 43210}}

\date{\today}
\maketitle

\begin{abstract}
We show analytically 
that the inverse kinetic inductance $L^{-1}$ of an overdamped 
junction array at low frequencies is proportional to the admittance 
of an inhomogeneous equivalent impedance network.  The $ij^{th}$ 
bond in this equivalent network has an inverse inductance 
$J_{ij}\cos(\theta_i^0-\theta_j^0-A_{ij})$, where $J_{ij}$ is the 
Josephson coupling energy of the $ij^{th}$ bond, $\theta_i^0$ is 
the ground-state phase of the grain $i$, and $A_{ij}$ is the usual 
magnetic phase factor. We use this theorem to calculate $L^{-1}$ 
for square arrays as large as $180\times 180$.  The calculated
$L^{-1}$ is in very good agreement with the low-temperature limit 
of the helicity modulus $\gamma$ calculated by conventional 
equilibrium Monte Carlo techniques.  However, the finite 
temperature structure of $\gamma$, as a function of magnetic field,
is \underline{sharper} than the zero-temperature $L^{-1}$, which 
shows surprisingly weak structure. In triangular arrays, the 
equilibrium calculation of $\gamma$ yields a series of peaks at 
frustrations $f = \frac{1}{2}(1-1/N)$, where $N$
is an integer $\geq 2$, consistent with experiment. 

\end{abstract}


\narrowtext

\maketitle

\newpage
\vspace{0.1in}

\section{Introduction.}

Superconducting arrays have been a subject of continuing interest 
for a number of years\cite{ref}.  Such arrays typically consist of 
a large collection of superconducting grains, of micron or 
submicron dimensions, arranged in a periodic structure.  They are 
embedded in a nonsuperconducting host which mediates a Josephson 
or proximity effect coupling between the grains.  The coupling 
generally produces a transition to a coherent state at a temperature
$T_{c}(B)$ which is well below the superconducting 
transition temperature $T_{c0}$ of the individual grains, and which 
strongly depends on the applied magnetic field $B$. There has been 
extensive research on the {\em thermodynamic} properties of these 
arrays, especially as a function of $B$ 
\cite{toboch,teitel,shih,shih2,choi,yosefin,berge}. More recently, 
a number of workers have extended these studies to the 
{\em dynamical} response of such arrays, especially their 
a.\ c.\ $IV$ characteristics\cite{steps} and vortex motions in the 
array\cite{vortex}.

In theoretical studies, the thermodynamic response of superconducting 
arrays is commonly characterized by the magnetic field and temperature 
dependent {\em helicity modulus},  $\gamma(B, T)$.  
$\gamma$ is defined as the stiffness of the array to a twist in the
phase of the superconducting order parameter\cite{teitel,shih,shih2}.
It is thus closely analogous to the spin wave stiffness in a magnetic 
material.  $\gamma$ can readily be expressed as an equilibrium 
quantity and evaluated as a thermodynamic average in the canonical 
ensemble, e.\ g., by Monte Carlo simulation. On the other hand, the 
{\em measurement} of $\gamma$ is typically a dynamical 
problem\cite{jean,theron}.  Such a measurement is based on the 
identification of $\gamma$ with the {\em inverse kinetic inductance}
$L^{-1}$, where $L$ is the effective inductance of the array.  $L$ 
can be measured at
temperatures below $T_c(B)$ by applying an appropriate a.\ c.\ current
and measuring the out-of-phase part of the a.\ c.\ voltage response.
Thus, in order to compare theory and experiment, one usually equates a 
thermodynamic quantity to a dynamical one.

In this paper, we describe a straightforward {\em dynamical} procedure
for calculating $\gamma$ in a superconducting array at temperature 
$T = 0$. The method starts from the coupled Josephson equations for 
the array in the limit of low frequency and small amplitude.  By 
carrying out an appropriate Taylor expansion in this limit, we show 
that $\gamma$ is equivalent to the
inverse inductance of a certain inhomogeneous impedance network.  The
impedances in this network are determined by the ground-state phases 
arrangement in the array.  Once the ground state is known, the 
effective impedance of the network can be calculated by standard 
numerical, or even analytical, techniques.

To show that this method is practical, we apply it to a square 
Josephson-coupled network subjected to an applied transverse magnetic
field.  The resulting $\gamma(T = 0)$ has surprisingly weak structure
as a function of magnetic field. Remarkably, the structure appears 
most conspicuously at {\em finite} temperature, because $T_c(B)$  is 
an extremely sensitive function of $B$.
To verify the consistency of this picture, we calculate $\gamma$ by
the standard equilibrium approach.  We show that, for all values of $B$
considered, it approaches the $T = 0$ value obtained by {\em dynamical}
techniques.

We have also used the equilibrium approach to calculate $\gamma(B,T)$ 
for a {\em triangular} lattice as a function of magnetic field.  The
dynamical approach is still applicable, in principle, to triangular 
lattices.  However, we have not used it in this case, because our  
algorithm for computing impedances does not always converge for 
triangular lattices.  Our calculated equilibrium $\gamma(B,T)$ has a 
very rich structure, as a function of $B$, which agrees in some detail
with that found in recent {\em dynamical} measurements by Th\'{e}ron 
{\it et al} \cite{theron}.  This agreement confirms, once again, that 
the static and dynamical quantities are basically equivalent, and also
shows that the array is well described by the standard frustrated XY 
model which underlies both the static and dynamic theories.

We turn now to the body of the paper.  Section II describes the 
formalism underlying our calculations.  In Section IIA, we prove the 
equivalence of $\gamma(T=0)$ to the effective impedance of an 
inhomogeneous impedance network. Section IIB reviews the standard 
equilibrium method for calculating $\gamma$.  Section III describes 
our numerical method for calculating both 
the effective impedance and the quantities needed for an equilibrium 
evaluation of $\gamma(T)$.  Our results for both square and triangular 
lattices are presented in Section IV.  A brief discussion follows in 
Section V.
    
\section{Formalism.}

Our basic geometry is shown in Fig.\ 1.  It consists of a periodic 
network of superconducting grains, each of which is Josephson coupled 
to its nearest neighbors.   We can calculate the inverse kinetic 
inductance $L^{-1}$, of this network by two different methods.  
In the first method, we use an identity, proven below, which shows 
that $L^{-1}$ is equivalent, within a proportionality constant, to the 
admittance of an inhomogeneous equivalent network.  This identity 
permits calculation of $L^{-1}$, or equivalently the helicity modulus 
$\gamma$, at temperature $T = 0$.  The second method is particularly 
appropriate at finite temperatures.   It uses the fact that $\gamma$ 
is an equilibrium 
thermodynamic quantity which can be calculated by standard Monte 
Carlo techniques.  We review this type of calculation
in Section IIB below.  
 
\subsection{Helicity Modulus at $T=0$:   Dynamical Evaluation.}

We consider an overdamped resistively-shunted Josephson junction (RSJ)
array in the presence of an external magnetic field ${\bf B}$ applied 
perpendicular to the array. The dynamical equations for the array at 
zero temperature take the form
\begin{eqnarray}
I_{ij} & = & \frac{V_{ij}}{R_{ij}}+ I_{c;ij}\sin(\theta_i-\theta_j
             -A_{ij}),\\
V_{ij} & = & \frac{\hbar}{2e}\frac{d}{dt}(\theta_i-\theta_j), \\
\sum_{j}I_{ij} & = & I_{i;ext}.
\end{eqnarray} 
Here $I_{ij}$ is the current from grain $i$ to grain $j$, $V_{ij} 
\equiv V_i-V_j$ is the voltage difference between grains $i$ and $j$, 
$I_{c;ij}$ and $R_{ij}$ are the critical current and shunt resistance
of the $ij^{th}$ junction, $\theta_i$ is the phase of the order 
parameter on the $i^{th}$ grain, and $I_{i;ext}$ is the external 
current fed into the $i^{th}$ grain. $A_{ij}$ is a magnetic phase 
factor defined by the relation
\begin{equation}
   A_{ij} = \frac{2\pi}{\Phi_0}\int_i^j{\bf A}\cdot{\bf dl}, 
\end{equation}
where ${\bf A}$ is the vector potential for the external magnetic 
field, and $\Phi_0=hc/2e$ is the flux quantum. 
Eq.\ (1) simply expresses the total current of each junction 
as the sum of a normal current through a shunt resistance and a 
Josephson supercurrent.  Eq.\ (2) is the Josephson relation connecting
the voltage drop across a junction to the phase difference.  Eq.\ (3) 
is just the Kirchhoff's Law, expressing the current conservation at 
the $i^{th}$ grain.  In the calculations below, we will assume that 
the vector potential is that of the externally applied magnetic field.
This is equivalent to assuming that the intergranular Josephson 
coupling is so weak that the screening magnetic 
fields can be neglected.

Let us assume that in the \underline{absence} of an external current, 
the $i^{th}$ phase has a value of $\theta_i^0$.  We assume that the
collection of phases $\{\theta_i^0\}$ represents a local energy 
minimum. One possible state of this kind is the true ground state of 
the array, but there may be other metastable states which also satisfy
this condition.  We now consider an a.\ c.\ external current 
$I_{i;ext} = $Re $[I_{i0}\exp(-i\omega t)]$, where $I_{i0}$ is an 
appropriate amplitude.  If the amplitude is not too large, the induced
voltages and phase changes will also be small, and we 
may assume that they also have frequency $\omega$.  Hence, we write the
phase on the $i^{th}$ grain as
\[  \theta_i(t) = \theta_i^0 + \delta\theta_i(t),  \]
We can now use these assumptions to linearize the equations of motion 
about 
the state $\{\theta_i^0\}$.  Combining Eqs.\ (1) -- (3), we can write
\[ \sum_j \left( I_{c;ij}\sin(\theta_i^0-\theta_j^0-A_{ij} 
    +\delta\theta_i-\delta\theta_j) 
    +\frac{\hbar(\dot{\delta\theta}_i-\dot{\delta\theta}_j)}
    {2eR_{ij}}\right) = I_{i;ext}.  \]
Expanding the sine function in a Taylor series to first order in
$\delta\theta_i-\delta\theta_j$, and using the fact that in the 
absence of external current, the state $\{\theta_i^0\}$ satisfies
\begin{equation}
\sum_jI_{c;ij}\sin(\theta_i^0-\theta_j^0-A_{ij}) = 0,
\end{equation}
we obtain the simple relation
\begin{equation}
  \sum_j \left[ (\delta\theta_i-\delta\theta_j)I_{c;ij}\cos(\theta_i^0-
  \theta_j^0-A_{ij})+\frac{\hbar(\dot{\delta\theta}_i
  -\dot{\delta\theta}_j)}
  {2eR_{ij}}\right]\nonumber = I_{i;ext}.
\end{equation}
Finally, we use the assumption that $\delta\theta_i$ is varying 
sinusoidally in time with frequency $\omega$ to rewrite this equation 
as
\begin{equation}
\sum_j\left[\frac{2ie}{\hbar\omega}I_{c;ij}\cos(\theta_i^0-\theta_j^0-
A_{ij}) +\frac{1}{R_{ij}}\right]V_{ij} = I_{i;ext},
\end{equation}
where we have used the relation
\begin{equation}
V_{ij} = \frac{\hbar}{2e}(\dot{\delta\theta}_i-\dot{\delta\theta}_j)
       = -i\frac{\hbar\omega}{2e}(\delta\theta_i-\delta\theta_j).
\end{equation}

We can interpret equations (7) in a very simple way.  Namely, they are 
identical to those for an 
\underline{inhomogeneous impedance network} responding to the 
driving a.\ c.\ current.  The $ij^{th}$ admittance is the sum of two
terms in parallel: a conductance $g_{ij} = 1/R_{ij}$, and a purely 
imaginary inductive element 
$[2ie/(\hbar\omega)]I_{c;ij}\cos(\theta_i^0-\theta_j^0-A_{ij})$.  
The elements of this network are all known if the unperturbed 
metastable state $\{\theta_i^0\}$ is known at the fields of 
interest.  Given the elements of the network, the 
\underline{effective network admittance} can be obtained by standard 
numerical techniques which have been developed for finding the 
admittances of inhomogeneous impedance networks\cite{kirk,frank}.

The results are particularly simple in the limit of low frequency. In
this case, the shunt resistances are irrelevant, and the response of 
the network is purely inductive.  The effective inductance of the 
network is then often 
known as the \underline{kinetic inductance}.
The present work permits this kinetic inductance to be
computed \underline{dynamically}, at least at $T=0$, by solving for 
the impedance of an effective impedance network.  A numerical 
technique for carrying out this calculation is summarized below.  

\subsection{Helicity Modulus at Finite Temperature: Equilibrium 
	    Evaluation.}

The kinetic inductance of an overdamped Josephson network can also be 
calculated as an \underline{equilibrium} quantity, by connecting it 
to the so-called helicity modulus of the network.  This connection has
been known for more than ten years\cite{teitel}, but for completeness 
we summarize this relation in the following paragraphs.

The equilibrium Hamiltonian of the Josephson network is given by
\begin{equation}
H = -\sum_{ij}E_{ij}\cos (\theta_i - \theta_j - A_{ij})
\end{equation}
where $E_{ij} = \hbar I_{c;ij}/(2e)$ is the coupling energy between 
grains $i$ and $j$, and $A_{ij}$ is given by Eq.\ (4).  In our 
calculations, we assume that $E_{ij}$ is the same for each junction. 
The equilibrium thermodynamic properties are given by the free energy
\begin{equation}
   {\cal F} = -k_BT\ln Q_N(T)
\end{equation} 
where the $N$-particle partition function $Q_N$ is given by the 
canonical integral
\begin{equation}
Q_N(T) = \int_0^{2\pi}\cdots\int_0^{2\pi}\Pi_i d\theta_i\exp(-H/k_BT).
\end{equation}
Similarly, the canonical average of any desired function 
${\cal O}(\theta_1, \cdots, \theta_N)$ is given by
\begin{equation}
   <{\cal O}> = \frac{1}{Q_N}\int\cdots\int \Pi_i d\theta_i{\cal O}
   (\theta_1, \cdots, \theta_N)\exp(-H/k_BT).
\end{equation}

The helicity modulus $\gamma_{\a\b}$ (or equivalently, the superfluid 
density) is defined as the free energy cost of imposing a twist in the
phase at the boundaries of the sample.  The principal elements of 
$\gamma_{\a\b}$ can be thought of as spin-wave stiffness constants 
appropriate to ``phase waves'' in this weakly coupled 
system.  Rather than imposing a twisted boundary condition and
calculating the resulting increase in free energy, however,
it is more convenient to
use periodic boundary conditions and calculate
$\gamma_{\a\b}$ as
\begin{equation}
  \gamma_{\a\b} = \left( \frac{\partial^2{\cal F}}{\partial 
  A_{\a}^{\prime}\partial A_{\b}^{\prime}} \right)_{\vec{A}^{\prime}=0}.
\end{equation}
Here $\vec{A}^{\prime}$ represents an added uniform vector potential 
(in addition to that which produces the applied magnetic field) which 
is included in the Hamiltonian in order to produce a twist.  The 
various second derivatives in Eq.\ (13) are readily computed for an
ordered or a disordered sample, with the result for, e.\ g., 
$\gamma_{xx}$:
\begin{eqnarray}
  N\gamma_{xx} & = & \left<\sum_{<ij>}E_{ij}x_{ij}^2
     \cos(\theta_i-\theta_j-A_{ij})\right>\nonumber \\ 
      & - & \frac{1}{k_BT}\left[\left<\left(\sum_{<ij>}E_{ij}x_{ij}
     \sin(\theta_i-\theta_j-A_{ij})\right)^2\right>\right]
	     \nonumber \\
      & + & \frac{1}{k_BT}\left[\left<\sum_{<ij>}E_{ij}x_{ij}
		     \sin(\theta_i-\theta_j-A_{ij})\right>^2\right],
\end{eqnarray}
where $x_{ij}=x_j-x_i$ and $x_i$ is the $x$ coordinate of the center of
grain $i$.  Similar expressions 
hold for the other components of
$\gamma$.  The above expressions are valid for both square and
triangular lattices.

We now connect $\gamma_{\a\b}$ to the inductive response of the array. 
To make this connection,
note that the {\em first} derivative of the free energy is the 
local current density ${\bf J}$, i.\ e.,
\[ \left(\frac{\partial {\cal F}}{\partial A_\a'}\right)_{\vec{A}'=0}
   = -J_\a/c.\]
Hence, 
\[ \gamma_{\a\b} = -\frac{1}{c}\left(\frac{\partial J_\a}{\partial 
   A_\b'} \right)_{\vec{A}'=0}.  \]
If $\vec{A}'$ is produced by an a.\ c.\ electric field $\vec{E}$ of 
frequency 
$\omega$, then $\vec{A}'=-ic\vec{E}/\omega$.  If we introduce the 
conductivity $\sigma_{\a\b}$ by
$\sigma_{\a\b} \equiv (\partial J_\a/\partial E_\b)_{\vec{E}=0}$, then 
it follows that $\gamma_{\a\b} = -i\omega\sigma_{\a\b}/c^2$. At very 
low frequencies, $\sigma_{\a\b}$ is of the purely inductive form 
$\sigma_{\a\b}=iL^{-1}_{\a\b}/\omega$, where 
$L_{\a\b}^{-1}=c^2\gamma_{\a\b}$.
Thus $\gamma_{\a\b}$ is indeed proportional to
low-frequency inductive response $L^{-1}_{\a\b}$ of the array.

In general, $\gamma_{\a\b}$ is a tensor, not a scalar.  In square or 
triangular arrays, however, the fourfold or sixfold rotational
symmetry implies that $\gamma_{\a\b}$ is, in fact, just a multiple of 
the unit tensor, i.\ e.\ $\gamma_{\a\b}(T) = \gamma(T)\delta_{\a\b}$.

\section{Method of Calculation.}

In all the calculations in this paper, we assume that the array is
homogeneous and ordered. 
That is, we assume that the coefficients $I_{c;ij}$ and $R_{ij}$ are 
independent of position and equal to $I_c$ and
$R$ respectively.   It follows that the coupling energies $E_{ij}$ 
are also position-independent and equal to
a constant which we denote simply as $E_J = \hbar I_c/(2e)$.

\subsection{Effective Impedance.}
 
For the impedance calculation, we inject a uniform a.\ c.\ current 
$I_{ext}$ into each grain at one boundary and extract the same current
from each grain in the opposite boundary.  To obtain the kinetic
inductance by the method of Section IIA, we need 
to compute the effective impedance of a certain equivalent network.
The inverse inductance of the $ij^{th}$ bond in this network is
$I_c\cos(\theta_i^0-\theta_j^0-A_{ij})$, where $\theta_i^0$ denotes 
the ground-state phase of the $i^{th}$ grain.
Thus, the problem divides into two parts: (i) finding the ground-state 
configuration $\{\theta_i^0\}$ in the absence of current; 
and (ii), given that configuration, computing the effective impedance 
of the network.  In practice, we cannot be assured that 
we have obtained the true ground state; in general, we will obtain a
{\em metastable} configuration which may be close to the ground state.

To obtain a metastable state near the ground state phase 
configuration, 
we carry out a standard Monte Carlo annealing\cite{halsey} at
fixed magnetic field, starting from a high temperature and cooling 
gradually.  The purpose of this annealing is to avoid, or at least to 
minimize, the trapping into metastable states which is a notorious 
consequence of dealing with frustrated systems, such as these arrays.  
In this part of the calculation, periodic boundary conditions are used
in all directions.  To check that we are indeed dealing with 
a near-ground-state configuration, we generally 
carry out more than 50 anneals at a given temperature.  The effective 
impedance is computed for the state with the lowest energy.

To calculate the effective impedance of the network, we need to solve 
the set of sparse linear equations (7).  In practice, at the 
low-frequency limit, the shunt resistance contribution can be 
discarded, and the trivial
common factor of $i/\omega$ divided out.
For both square and triangular arrays, we then use the Gauss-Seidel 
iterative method to solve for the impedance of 
networks as large as $180 \times 180$. This iterative procedure always 
converges in our square arrays calculations. In our
triangular array calculations, the iterative approach sometimes fails
to converge, possibly because the ground state factors
$\cos(\theta_i^0-\theta_j^0-A_{ij})$ can be of both signs for 
triangular arrays.  Thus, we will present results for 
triangular arrays
using only the equilibrium method described above.

\subsection{Helicity Modulus at Finite Temperature.}

To calculate the helicity modulus $\gamma_{\a\b}$ at finite 
temperature, we use
the equilibrium Monte Carlo method within the standard Metropolis 
algorithm\cite{binder} for both square and triangular arrays with 
periodic boundary conditions. In order to minimize 
the effects of metastable states, 
we cooled our system from high temperatures at each magnetic field, 
typically starting from $1.5E_J/k_B$ and cooling to 
$0.5E_J/k_B$ in steps of $0.1E_J/k_B$, followed by an
additional cooling down to $0.2E_J/k_B$ in steps of
$0.05E_J/k_B$.   Following this cooling, the required averages
were obtained by making 60000 passes through 
the entire lattices, with the first 10000 passes discarded.
We carry out the thermal average by including only every tenth pass
through the entire array, in order to
minimize the strong statistical correlation between successive 
configurations.  The reported results are averages of 
$\gamma = (\gamma_{xx} + \gamma_{yy}) / 2$ and are generally produced 
by averaging over four to eight independent runs.

\section{Results.}

\subsection{Square Arrays.}

Fig.\ 2 shows the helicity modulus $\gamma$ plotted as a function of
frustration values $f$ at four different temperatures in square arrays.
The array sizes are either $48\times 48$ or $50\times 50$ in the 
finite-temperature calculations, which are carried out by the 
equilibrium technique of Section IIB.  In the $T=0$ calculations, we
use arrays from $120\times 120$ to $180\times 180$ and the analog 
technique of Section IIA. The frustration $f \equiv \Phi/\Phi_0$,
where $\Phi$ is the flux per plaquette. Note that as temperature 
decreases, these Monte Carlo 
calculations converge remarkably well to the $T = 0$ impedance results.
This supports the validity of our method for 
calculating the kinetic inductance at temperature $T = 0$.

For $f = 0$ and $f = 1/2$, the melting temperature is known to equal 
approximately $0.95E_J/k_B$ and $0.45E_J/k_B$ 
respectively\cite{toboch,teitel}.
For $f=p/q$, with $p$ and $q$ mutually prime integers, as $q$ 
increases, the melting temperature, estimated as the point
where $\gamma$ goes to zero, decreases very quickly,
consistent with previous work\cite{teitel}. 
At $f=5/12$, for example, the melting temperature is probably less than
$0.1E_J/k_B$ (the vanishing of $\gamma$ at this value of $f$ is 
considerably broadened by finite size effects). 

A remarkable feature of the results shown in Fig.\ 2 is that the plot 
of $\gamma(f)$ becomes \underline{sharper} as the temperature is 
increased.
This sharpening with increasing temperature is seemingly contrary to 
intuition, but in fact is easily explained.  As noted above,
the melting temperature of the vortex lattice at a frustration 
$f = p/q$ decreases with increasing $q$ (assuming $p$ and $q$ mutually
prime integers).  Thus, at extremely low temperatures, the vortex 
lattice is stable for a wider range of $f$ values,
but at slightly higher temperature, it has melted at all except a few
values of $f$ corresponding to relatively small values of $q$.  The 
corresponding $\gamma(f)$ shows sharp peaks only at a few such values 
of $f$.  (As seen in Fig.\ 2, the sharpest features at a temperature 
of $0.1E_J/k_B$ occur at $f = 1/5, 1/3, 2/5$, and $1/2$.)
This explains why the \underline{zero-temperature} $\gamma(f)$ is  
surprisingly featureless: for all values of $f$,
the vortex lattice is apparently frozen at $T = 0$. 

It should be emphasized that the $T = 0$ calculations are carried out
using ``ground states'' obtained by Monte Carlo annealing from a 
finite temperature.  As pointed out in the previous section, such 
annealing is not guaranteed to produce the true ground state, but 
only a metastable state 
close to the ground state.  We have checked for metastability by 
calculating $\gamma$ at $T=0$ using several such annealed states.  The 
resulting $\gamma$'s are generally quite close to that obtained in the
state of lowest energy.  The $\gamma$'s plotted in Fig.\ 2 always
correspond to this lowest energy state.

\subsection{Triangular Arrays.}

Fig.\ 3 shows the calculated helicity modulus $\gamma$, plotted as a 
function of the frustration value $f$ in a triangular array at four 
different temperatures. As in a square array,  when $f = p/q$, where 
$p$ and $q$ are mutually prime integers, the melting temperature tends
to decrease strikingly with increasing $q$.  Indeed, when $q$  
is sufficiently large (typically, when $q \ge 8$), the melting 
temperature of the array is less than $0.1E_J/k_B$).  As in the square
array, there are a series of sharp peaks in $\gamma(f,T)$ plotted as 
a function of $f$ at fixed $T$.  Some of this structure has been 
previously studied\cite{shih}, but the present work represents a 
considerably more detailed investigation.  We have found clearly 
distinguishable peaks in $\gamma$ at frustration values 
$f=\frac{1}{2}, \frac{1}{3}, \frac{1}{4}, \frac{2}{5}, \frac{3}{7}, 
\frac{3}{8}, \cdots$. Indeed, we have obtained peaks at every value 
of $f$ between 1/3 and 1/2 
satisfying the relation $f=\frac{1}{2}(1-1/N)$, with N as large as 8.
All such peaks have been observed in recent 
experiments by Th\'{e}ron {\it et al}\cite{theron}.
Since we are able to obtain such peaks here, we conclude that the 
experimental arrays are adequately described by a frustrated XY model,
as assumed here.  Of course, in the calculations, we are unable to 
vary $f$ \underline{continuously} as is possible in the experimental 
geometry. Thus, we are able to identify ``peaks'' only by considering 
nearby values of $f$ not satisfying the relation 
$f= \frac{1}{2}(1-1/N)$, and showing 
that these have smaller values of $\gamma$ than the $f$ values in the 
chosen series. We have shown a few such values in Fig.\ 4.  As one 
illustration, the value of $\gamma$  at $f=3/8$ are larger than 
those at $f=17/48$ and $f=19/48$. 

Th\'{e}ron {\it et al}\cite{theron} account for 
the peaks at $f=\frac{1}{2}(1-1/N)$
on the basis of a superlattice of
vacancies in a background of the well-known checkerboard vortex
ground state which forms at $f = \frac{1}{2}$.  In their model,
a change in the nature of the ground state is predicted at a
critical frustration $f_c\approx 0.468$.  When 
$f_c < f < \frac{1}{2}$, a 2D vacancy superlattice is found to
be energetically favored, while for $1/3 < f < f_c$,
a vortex superlattice consisting of 1D parallel stripes is preferred.

In order to check this conjecture in our simulation, we have 
calculated the low-temperature vortex configuration of the vortex 
lattice at various values of $f$.  In a Josephson junction array 
with frustration $f$, the vortex
number $n$ of each plaquette is defined as:
\[ \sum_{plaquette} (\theta_i - \theta_j - A_{ij}) + 2\pi f= 2\pi n \]
where the gauge invariant phase difference for each junction 
$(\theta_i-\theta_j - A_{ij})$ is in the range $(-\pi, \pi]$ and the 
summation is taken in the counterclockwise direction.  

Some of our calculated vortex configurations are shown in Fig.\ 4.
Fig.\ 4 (a) shows the configuration at $f = 3/8 = 0.375$ at
temperature $T=0.01E_J/k_B$. It is easy to see there is indeed a
striped domain structure in the vortex lattice, although the domain 
wall configuration is somewhat different from that predicted 
by \cite{theron}. For lattices such that $f>f_c$, it is necessary to 
carry out the calculations for relatively large lattices in order to 
distinguish the results from the $f = 1/2$ case.  It proves very 
difficult to observe the predicted vacancy lattice
structure.  Fig.\ 4 (b) shows a representative low-temperature vortex 
configuration for an $f$ value in this range, namely
$f = 17/36 = 0.472$ at a temperature $T=0.01E_J/k_B$. It is difficult 
to categorize the vortex lattice into any of the domain 
patterns suggested by \cite{theron}.  Nevertheless, 
we do obtain nearly all the peaks found experimentally, at most of the
fields studied.  We conclude that, while the 
experimental peaks do correspond to particularly stable vortex 
structures, there are many structures at these frustrations
which are nearly equally stable, and which could all give rise to the 
observed peaks. 

Finally, we have checked for hysteresis at $f = 1/6$ in order to test 
whether the melting transition at that field is first-order. In
$18\times 18$ triangular arrays, we have used a Monte Carlo annealing 
method first to cool the array from a temperature $1.5E_J/k_B$ to 
$0.1E_J/k_B$ gradually, and then to reheat it.  We find a clear 
hysteresis in the energy in the temperature range between 
$0.25E_J/k_B$ and $0.32E_J/k_B$ [cf. Fig. 5], suggestive of a 
first-order transition. This result is consistent with expectations 
based on \underline{three-dimensional} calculations on stacked 
triangular lattices by Hetzel {\it et al}\cite{hetzel}, at $f = 1/6$.  

\section{Conclusion.}

In this paper, we have developed a new method to calculate the kinetic
inductance of a Josephson junction array at zero temperature. This 
method maps a junction array onto a linear complex impedance network.
The mapping permits us to calculate the zero-temperature kinetic 
inductance using much larger arrays than previously possible, provided
that the ground state can be obtained by other means. The kinetic 
inductance calculated by this method agrees well with Monte Carlo 
calculations of the helicity modulus, in the low temperature 
limit.  In both square and triangular arrays, 
the helicity modulus shows a series of
peaks as a function of the frustration $f$.  In triangular arrays,
these peaks occur at $f=\frac{1}{2}(1-1/N)$, where $N$ is an integer 
$\geq 2$, in agreement with experiment.

The dynamical method discussed here has other potential applications.
For example, it can be extended to treat arrays at finite frequencies,
i.\ e., to study the variation of the effective admittance with 
frequency, simply by retaining the shunt resistances in Eq.\ (7).   
It can also be used with little change to treat underdamped arrays.  
Finally, it may be possible to extend this approach to finite 
temperatures, by including a 
Langevin noise term in the equations of motion for the phases.

\section{Acknowledgments.}

We are grateful to R. \v{S}\'{a}\v{s}ik and X. Zhang for valuable 
conversations.  This work has been supported by NSF grant DMR90-20994,
and by the Midwest Superconductivity Consortium at Purdue University 
through D. O. E. 
grant DE-FG90-02-ER-45427.  Calculations were carried out, in part, 
on the CRAY Y-MP 8/8-64 of the Ohio Supercomputer Center.

\newpage
\begin{center}
	{\bf Figure Captions}
\end{center}

\begin{enumerate}
  \item Schematic diagrams of (a) an $8\times 8$ square and (b) an 
        8 $\times$ 8 triangular Josephson junction array.  Each 
	intersection represents a superconducting grain, which is 
	connected to its nearest neighbors by Josephson coupling.  
	In the dynamical inductance calculation, the external current 
	is injected into each grain at one boundary and extracted 
	from the other side boundary; free boundary conditions 
        are used in the direction of current injection, while periodic 
        boundary conditions are used in the transverse direction.
        In the Monte Carlo calculations,
	periodic boundary conditions are applied in both directions.

  \item Calculated helicity modulus versus frustration $f$ at four
	different temperatures in square arrays. Lines merely serve to 
	guide the eye.   For clarity, the curves for $T = 0, 
	0.01E_J/k_B$ and $0.05E_J/k_B$ are shifted vertically by 0.3,
	0.2, and 0.1 units respectively. The zero temperature result 
	is obtained from a 
	dynamical calculation of the inverse array inductance, using
        array sizes ranging from $120\times 120$ to $180\times 180$. 
        The finite temperature results are obtained by Monte Carlo 
	methods in $50\times 50$ arrays for $f=1/5, 2/5$, and in 
	$48\times 48$ arrays 
        at all other values of $f$ shown.

  \item Calculated helicity modulus versus frustration $f$ at four
	different temperatures in triangular arrays, as obtained by 
	Monte Carlo simulations. Lines merely serve to guide the eye.
  	For clarity, we have vertically shifted the plots for
	$T = 0.01E_J/k_B, 0.05E_J/k_B$ and $0.1E_J/k_B$ by
 	by 0.3, 0.2, and 0.1 units respectively.
	The array sizes are $56 \times 56$ for $f=3/7$; 
        $50 \times 50$ for $f=1/5$ and $2/5$; and $48 \times 48$
	for all other $f$ values shown.

  \item  Vortex configurations for triangular arrays as obtained by 
	Monte Carlo annealing at temperature $T=0.01E_J/k_B$ at two 
	different values of $f$: (a) $f=3/8$, $32\times 32$ array (the
 	striped pattern can be seen by viewing the array from lower 
	right to upper left); (b) $f=17/36$, $36\times 36$ array.  
        Dark and light 
        plaquettes denote the presence and absence of vortices.

  \item Energy per grain in $18 \times 18$ triangular arrays at 
	$f = 1/6$, as calculated on cooling (solid disks) and on 
	heating (open disks). 
        The displayed results represent averages over five independent 
        runs; lines serve to guide the eye.

\end{enumerate}

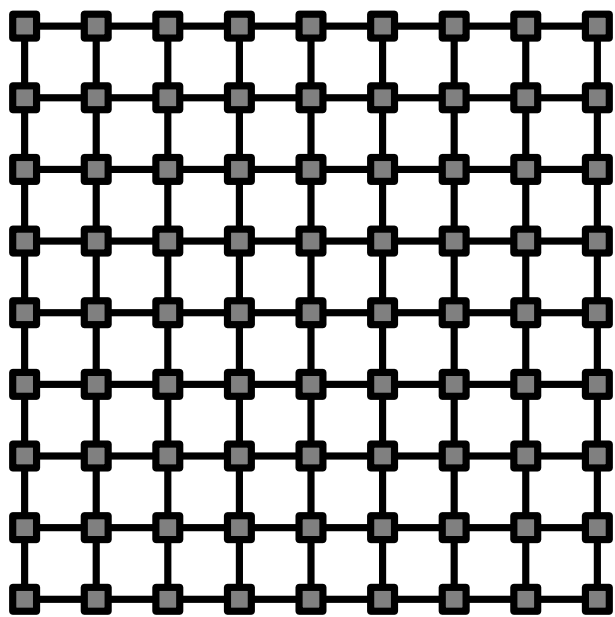
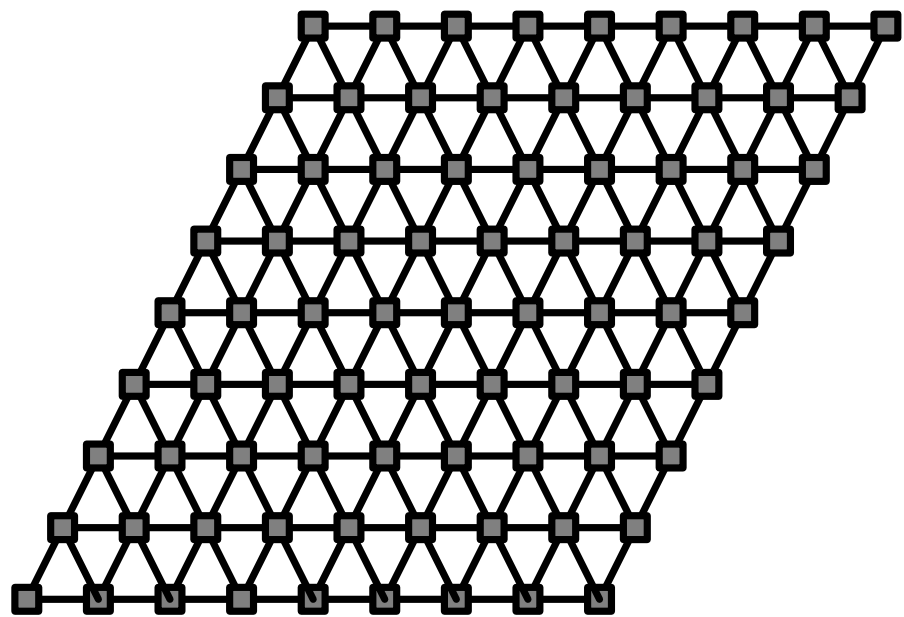
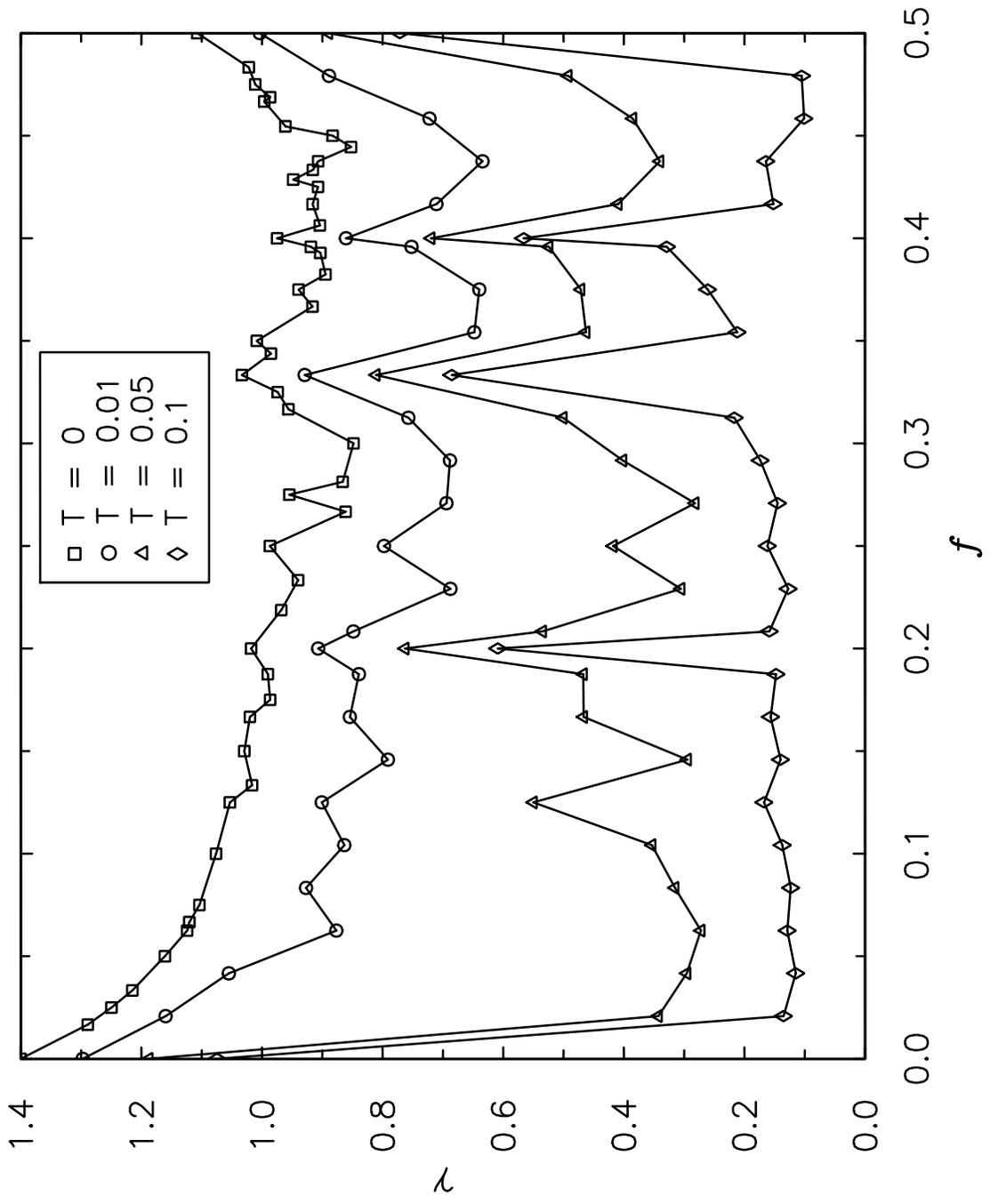
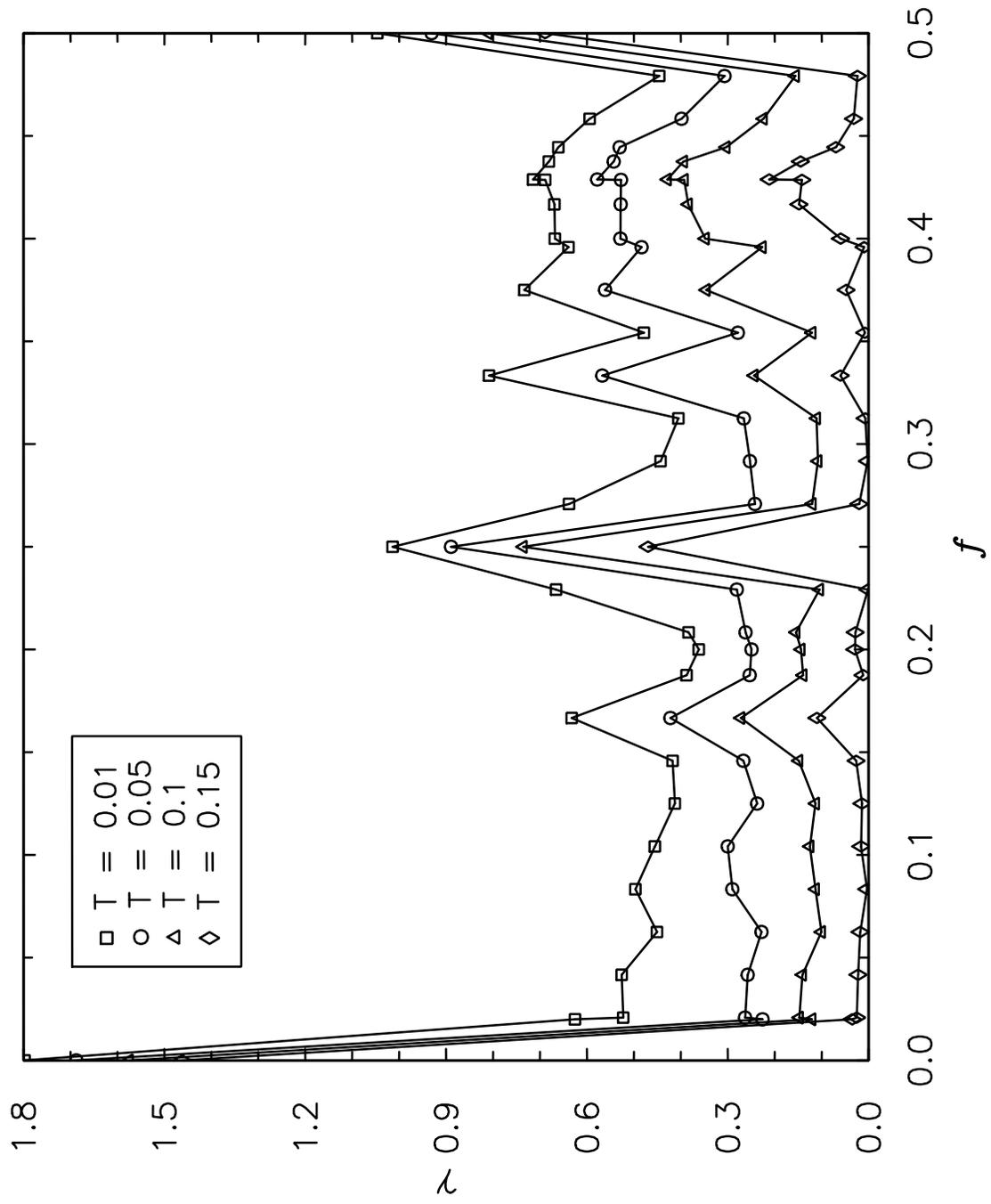
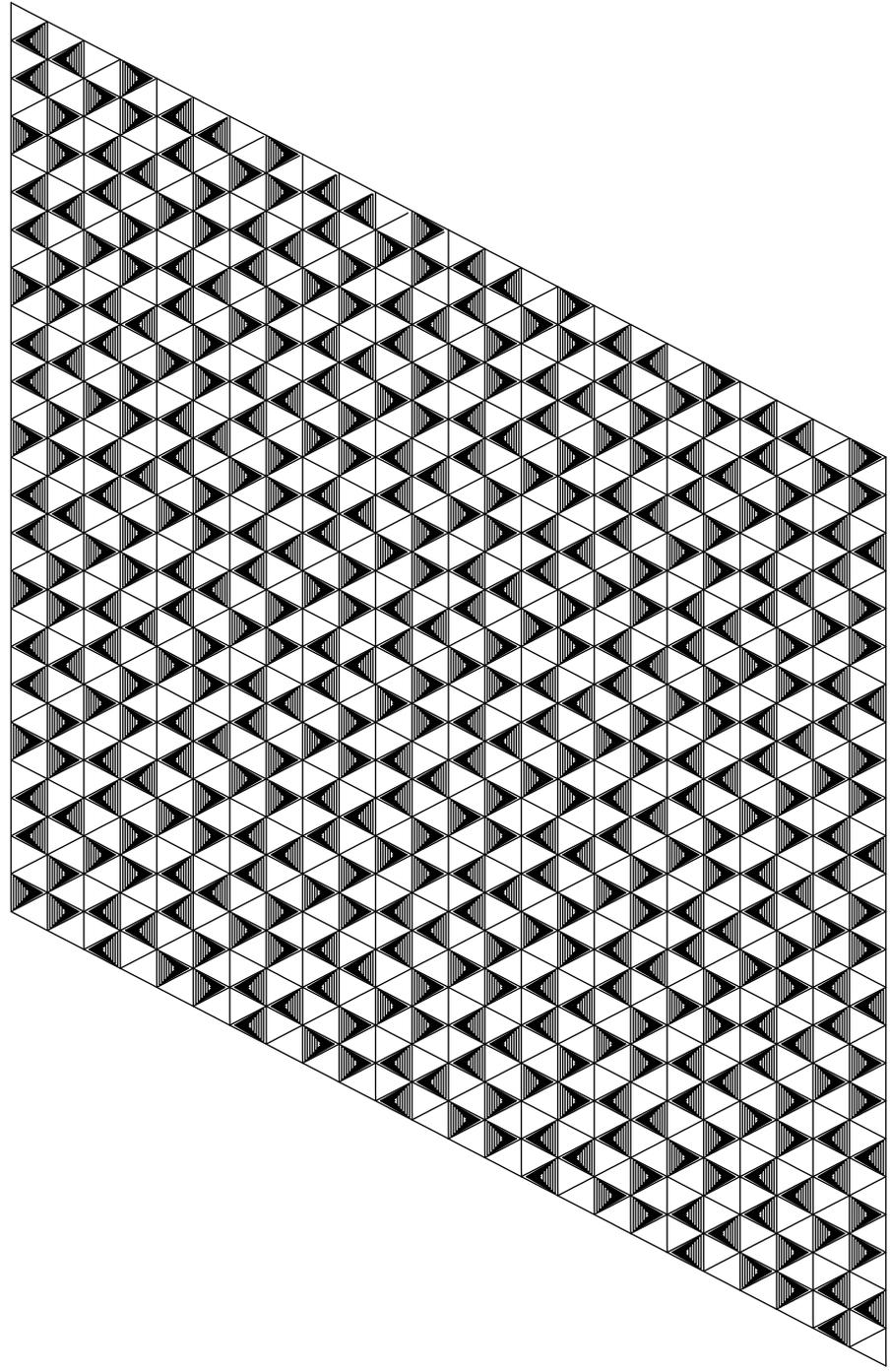
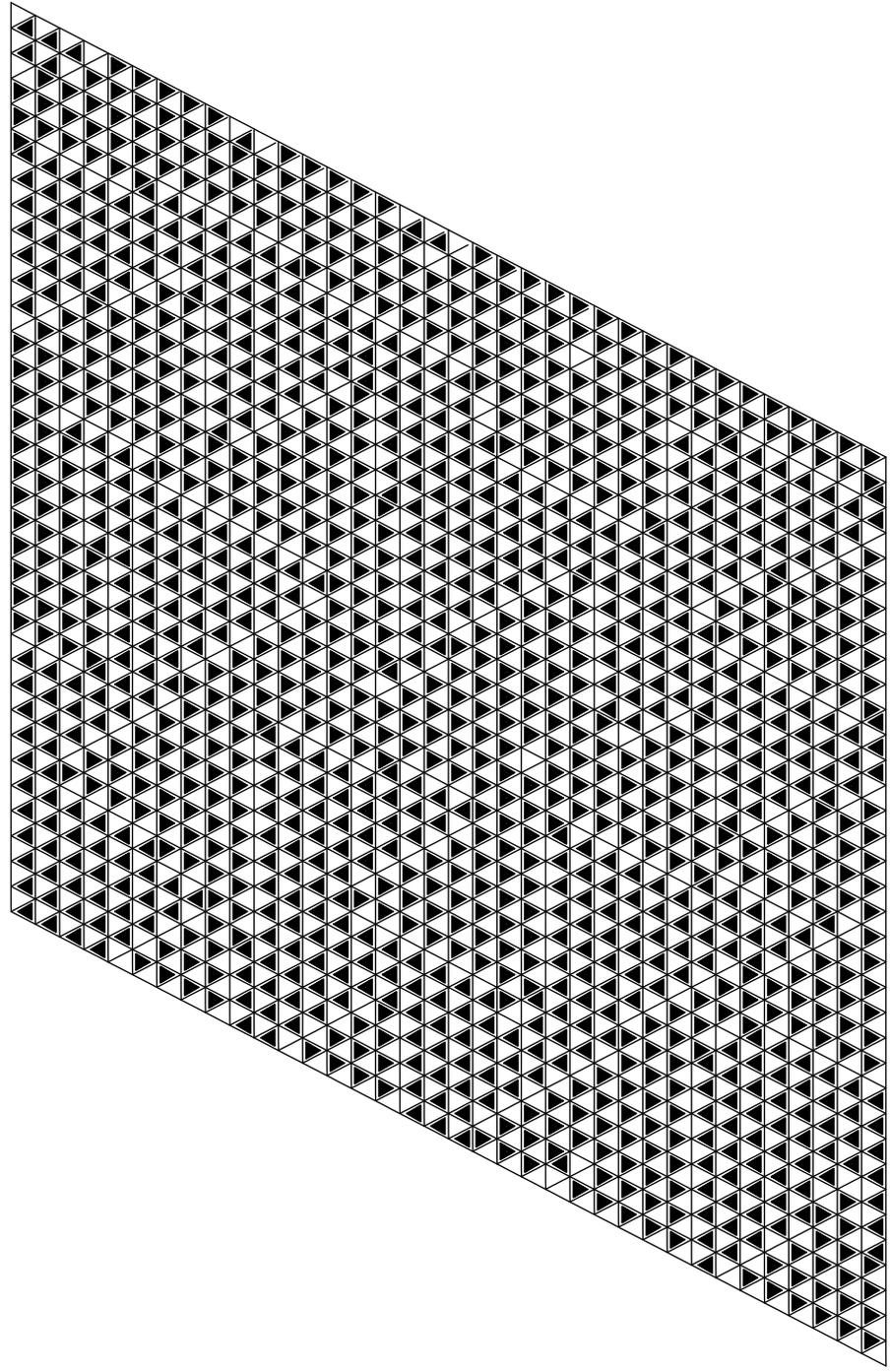
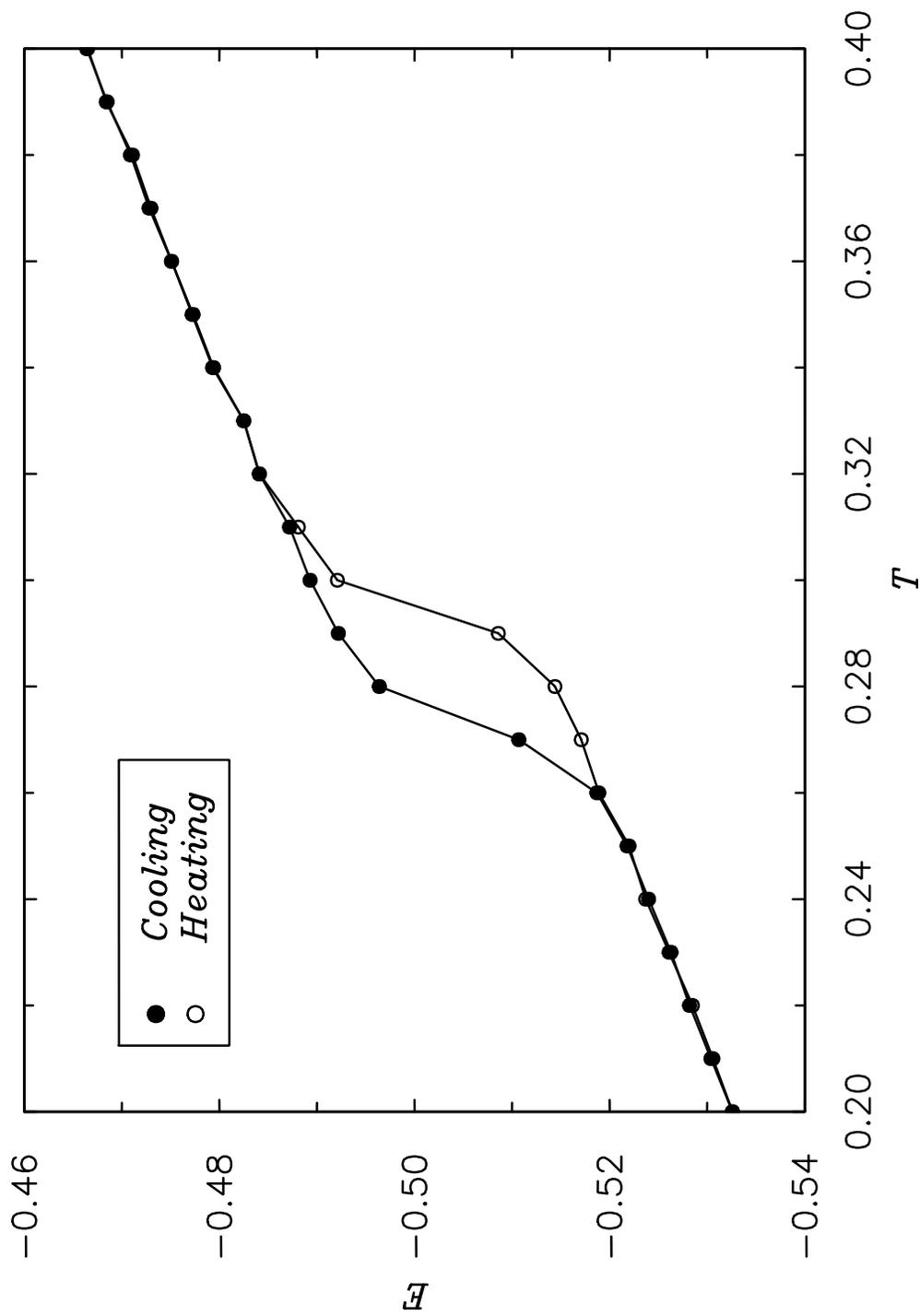
\end{document}